\documentclass[11pt,twoside,dvips]{article}
\usepackage{verbatim}
\usepackage{amssymb}
\usepackage{amsfonts}
\usepackage{amsmath}
\usepackage{pifont}
\usepackage[spanish,english]{babel}
\usepackage[pdftex]{graphicx}
\usepackage{verbatim}
\usepackage{epsfig}
\usepackage{bm}
\usepackage{longtable}
\usepackage{fancyhdr}
\usepackage{multirow}

\begin{document}

\fancypagestyle{plain}{%
\fancyhf{}%
\fancyhead[LO, RE]{XXXVIII International Symposium on Physics in Collision, \\ Bogot\'a, Colombia, 11-15 September 2018}}

\fancyhead{}%
\fancyhead[LO, RE]{XXXVIII International Symposium on Physics in Collision, \\ Bogot\'a, Colombia, 11-15 September 2018}

\title{Neutrino Physics with Reactors}
\author{Bed\v{r}ich Roskovec$\thanks{%
e-mail: beroskovec@uc.cl}$
 \\ Instituto de F\'{i}sica, Pontificia Universidad Cat\'{o}lica de Chile \\ Avda.~Vicu\~{n}a Mackenna 4860, Santiago, CHILE}
\date{\today}
\maketitle

\begin{abstract}
Rector neutrinos have been a tool to investigate neutrino properties for more than 60~years. The reactor neutrino flux was measured throughout 80s-90s. In the 2000s, reactor neutrino oscillations at large baselines were observed by the KamLAND experiment and later in 2012 at short baselines by the Daya Bay, Double Chooz and RENO experiments. Reactor neutrino experiments have significantly contributed to our current knowledge of oscillation parameters. The detector technology has been majorly improved over decades and we have entered era of precise measurements. The recent absolute measurement reactor neutrino flux and spectral shape revealed deviations from the prediction model, known as reactor antineutrino flux and spectrum shape anomalies. In this article, we review the latest development in short baseline reactor experiments and we discuss observed anomalies.  
\end{abstract}

\section{Introduction}
Nuclear reactors are a powerful source of pure electron antineutrinos with energy up to $\sim$10~MeV. They have been used to study neutrino properties for more than 60~years. The history of reactor neutrino measurements started with the discovery of electron neutrino by Cowan and Reines in the famous Savannah River experiment \cite{reines1}. With the spread of nuclear reactors, the 80s-90s saw the advent of many short baseline measurements of reactor neutrino flux and spectrum \cite{80s1, 80s2, 80s3, 80s4, 80s5}. Reactor neutrino experiments contributed also to the establishment of neutrino oscillations starting with KamLAND's measurement of $\Delta m^2_{21}$-driven neutrino oscillations at long baselines \cite{Kamland}.

The current generation of large scale liquid scintillator reactor neutrino experiments, Daya Bay, Double Chooz and RENO, were designed to study $\bar\nu_e$ disappearance at short baselines $\sim$1~km. These experiments use near and far detector(s) to cancel out correlated systematics in the oscillation measurements such as the reactor flux uncertainty and thus have significantly improved precision over single detector experiments. The big leap in precision led in 2012 to the discovery of reactor neutrino oscillations at short baselines, implying a  non-zero value of $\theta_{13}$ mixing angle \cite{DYB_first, RENO_first, DC_first}. The $\theta_{13}$ mixing angle was the last unknown in three-flavor neutrino mixing framework and its large value opened the road towards the measurement of combined charge-parity symmetry violation in the lepton sector. 

The unprecedentedly large statistics of antineutrinos detected in current generation experiments, such as Daya Bay's nearly 4 million events  \cite{DYB_latest}, allows to study in detail the reactor neutrino flux and energy spectrum shape. The measurements of the absolute reactor neutrino flux exhibit about 5\% lower overall rate compared to the reevaluated prediction from 2011 \cite{Huber, Mueller}, which is known as reactor antineutrino anomaly. The source of the anomaly is unknown. The existence of sterile neutrino(s) with mass $\sim$1~eV was proposed as a possible explanation. Oscillation from the electron to the sterile flavor would result in the observed deficit. Nevertheless, there might be issues with the prediction as suggested in Daya Bay's study of the evolution of reactor neutrino flux and spectral shape with nuclear fuel composition. The yield particularly for $^{235}\text{U}$ is different from the measurement \cite{DYB_evolution}.

In addition to the discrepancy in the absolute flux, Daya Bay, Double Chooz, RENO \cite{DYB_spectrum, bump_DC, RENO_bump}, as well as other short baseline experiments such as NEOS \cite{NEOS_bump}, observed a statically significant excess over the prediction in the $\bar\nu_e$ energy spectrum at the energy range of $5-7\text{ MeV}$. The size of this `bump' structure was demonstrated to be related to the reactor neutrino production process, thus pointing to an inaccurate prediction.

In this article, we give a brief introduction to reactor neutrino production and two basic prediction methods. We mention $\bar\nu_e$ detection via inverse beta decay (IBD) and briefly discuss IBD backgrounds. We describe the design of current generation of large-scale liquid scintillator experiments Daya Bay, Double Chooz and RENO, highlight their similarities and differences and summarize results for neutrino oscillation measurements in short baselines based on relative comparison between near and far detectors. Finally, we focus on absolute measurements where deviations from prediction are observed in absolute reactor neutrino flux and energy spectrum shape. We provide some proposed explanations focusing mainly on possible prediction issues. We conclude with brief outlook of reactor neutrino experiments.

\section{Reactor Neutrinos}
Fission of heavy nuclei occurs in nuclear reactors. Commercial reactor cores with up to few TW thermal power typically use low enriched uranium (LEU) fuel where the fissions of four isotopes, namely $^{235}$U, $^{238}$U, $^{239}$Pu and $^{241}$Pu, account for 99.9\% of emitted antineutrinos. Their relative contribution changes significantly during the fuel cycle, which last typically several months.  $^{235}$U is mostly burned in the beginning of the cycle, while plutonium isotopes are building up, with $^{239}$Pu becoming the largest contributor at the end. On the other hand research reactors typically use highly enriched uranium (HEU), where almost exclusively $^{235}$U is burned thoughout their whole fuel cycle. These reactors are not as powerful as the commercial ones.

The fission products are neutron rich isotopes, which undergo a series of beta decays until they reach stability. The reaction can be in general expressed as $^A_NX\rightarrow^A_{N-1}Y+e^-+\bar\nu_e$. Beta decays are a source of pure electron antineutrinos. There are $\sim$$6\:\bar\nu_e\text{'s}$ produced per fission, which corresponds to $\sim$$2\times10^{20}\:\bar\nu_e/\text{s/GW}_\text{th}$. Nuclear reactors are thus the most powerful man-made sources of electron antineutrinos. The energy of reactor neutrinos spans up to about 10~MeV.

\subsection{Reactor Neutrino Flux and Spectrum Prediction}
There are in general two complementary methods to predict the reactor neutrino flux and spectrum. The first one is summation method, where contributions of all decay branches of all possible fission isotopes is summed with appropriate weights in order to obtain the aggregate antineutrino spectrum. This method however involves large uncertainties for some important fission products and suffers from the disadvantage that not all branching ratios are known. As a result, it has larger total uncertainty compared to conversion method. The latter method turns the electron spectra for $^{235}$U, $^{239}$Pu and $^{241}$Pu, measured at ILL \cite{ILL1, ILL2, ILL3} in the 80s, into antineutrino energy spectra. Recently, electron spectrum for $^{238}$U was measured for \cite{U238} as well. The electron spectra, intrinsically summed over all possible decays, are fitted by number of virtual branches, whose shape takes into account several aspects such as forbidden decays, etc. The recent reevaluation of the conversion method for $^{235}$U, $^{239}$Pu and $^{241}$Pu isotopes \cite{Huber} with the summation calculation for $^{238}$U~\cite{Mueller}, usually referred to Huber+Mueller model, currently stands as the leading prediction model due to lower claimed uncertainties. The reevaluation has increased the predicted flux, resulting in the reactor antineutrino anomaly. 

\subsection{Detection via Inverse Beta Decay}
The reaction used for reactor neutrino detection in current liquid scintillator experiments is inverse beta decay (IBD) $\bar\nu_e+p\rightarrow e^++n$. Prompt positron's energy losses and its annihilation forms a correlated pari with subsequent delayed neutron capture. This coincidence signature allows to powerfully suppress the backgrounds. Moreover, the prompt energy is linked to the initial antineutrino energy through $E_{\bar\nu_e}\simeq E_{prompt}+0.78\text{ MeV}$, with a threshold $E_{\bar\nu_e}=1.806\text{ MeV}$. The neutron capture takes place on elements with high neutron capture cross-section, such as the atoms of hydrogen naturally present in liquid scintillator or special elements that are added to improve the delayed signal characteristics. For example, liquid scintillator of Daya Bay, Double Chooz and RENO was doped with gadolinium in order to achieve a distinct dalayed signal of 8~MeV.

These experiments, discussed in detail later, achieved very low background amounting to only 2\%, 5\% and 5\% of their far detecor(s) IBD samples, respectively. The largest source is accidental coincidence of two otherwise uncorrelated signals, which happens to pass the selection criteria. The Double Chooz experiment further employed artificial neural network to achieve a remarkable suppression of this background. This was needed since they have included among others neutron capture on hydrogen in their analysis, which experience significantly higher background rate than neutron capture on gadolinium. Backgrounds with the largest uncertainties are linked to cosmic-ray muons. These can produced long-lived unstable isotopes of $^9$Li and $^8$He, whose decays irreproducibly mimic IBD signature, or produce fast neutrons which can too due to possible scatter and subsequent capture look like IBD. Reducing these backgrounds is the primary reason these experiments are built deep underground, where the cosmic-ray muon flux is suppressed.

\section{Short Baseline Experiments: Daya Bay, Double Chooz, RENO}
The current generation of large liquid scintillator experiments, Daya Bay, Double Chooz, RENO, was designed with the aim to measure the last mixing angle to be determined, $\theta_{13}$. The previous measurement done by CHOOZ \cite{CHOOZ} was able to set only upper limit due mainly to uncertainties in the flux prediction and the absolute detection efficiency. This was addressed in the following experiments employing functionally identical near and far detector(s). Originally proposed in \cite{near_far_proposal}, the near detector provides a benchmark measurement for the far detector, which ideally sits in an optimized distance to experience the strongest possible oscillation effect. Relative comparison of far/near measurements in the ideal case, one reactor core and single near and far detectors, cancels out correlated uncertainties leaving only uncorrelated, typically much lower, in play. Having multiple reactors and detectors leaves some residual effect of the correlated uncertainties.

Table~\ref{tab:det_parameters} summarizes the main parameters for the three experiments. Daya Bay has largest reactor power, target mass and overburden. It is also the only experiment among these three that deploys more than one antineutrino detector at the same experimental site, allowing to direct comparison and test of identical performance. Double Chooz has near end far detector located in a so-called iso-flux configuration, when both near and far detectors have the same relative contribution from two reactors. This achieves by doing so the ideal case of correlated reactor flux uncertainty cancelation.

\begin{table}
\small
\begin{tabular}{|c|c|c|c|c|} \hline
Experiment	&	\begin{tabular}{c}Reactor Power \\ $\left[\text{GW}_{th}\right]$\end{tabular} 	&\begin{tabular}{c}Detector GdLS Mass [t]\\	Far (Near)	\end{tabular} 	& \begin{tabular}{c}Distance [m] \\Far (Near) \end{tabular} & \begin{tabular}{c}Overburdern [\text{mwe}]\\ Far (Near)\end{tabular}  \\ \hline
Daya Bay		& 6$\times2.9$	& \begin{tabular}{c}$4\times20$ \\ ($2\times2\times20$) \end{tabular}	& \begin{tabular}{c}1650 \\ (365,490) \end{tabular}	& \begin{tabular}{c}860 \\ (250) \end{tabular} 	\\\hline
Double Chooz	& 2$\times4.25$& \begin{tabular}{c}8 \\ (8) \end{tabular}							& \begin{tabular}{c}1050 \\ (400) \end{tabular}	&\begin{tabular}{c}300 \\ (120) \end{tabular} \\\hline
RENO		& 6$\times2.8$	& \begin{tabular}{c}16 \\ (16) \end{tabular}							& \begin{tabular}{c}1380 \\ (290) \end{tabular}	& \begin{tabular}{c}450 \\ (120) \end{tabular}	\\\hline
\end{tabular}
\caption{\label{tab:det_parameters}Summary of three major experiments parameters: Reactor power, mass of near and far detectors, their distances from reactor cores and overburden of experimental sites.}
\end{table}

All three experiments use cylindrical three-zone Antineutrino Detectors (ADs). The innermost zone is filled with gadolinium-doped liquid scintillator. The middle zone is filled with pure liquid scintillator and catches $\gamma$'s escaping from the central zone. The outermost zone contains photomultipliers (PMTs) for scintillation light detection.  This zone is filled with transparent non-scintillating mineral oil which shields against natural radioactivity $\gamma$'s to enter sensitive volume. All zones are separated with transparent acrylic vessel and are surrounded by a stainless steel vessel. ADs are surrounded by a muon tagging system. In Daya Bay and RENO, it is an ultra-pure water pool Cherenkov detector, while in case of Double Chooz it is another liquid scintillator layer, in all cases instrumented with PMTs. Double Chooz uses an additional layer of plastic scintillators over the detectors, while Daya Bay has four layers of resistive plate chambers. 

\section{Reactor Neutrino Oscillations at Short Baselines}
\subsection{Brief Introduction to Neutrino Oscillations}
Neutrino flavor oscillations are nowadays a well established phenomenon that has been observed in a great variety of experiments. Neutrinos are produced and detected as flavor states $\nu_e$, $\nu_\mu$, $\nu_\tau$, which are eigenstates of the weak interactions. However, these states are not mass eigenstates, but their superposition. The transformation between flavor and mass states $\nu_1$, $\nu_2$, $\nu_3$ is expressed by the Pontecorvo-Maki-Nakagawa-Sakata $3\times3$ unitary matrix. The parameters of this matrix that are relevant for neutrino oscillations are three mixing angles $\theta_{12}$, $\theta_{23}$, $\theta_{13}$ and one CP-violation phase $\delta$. Relative phase difference occurs in the neutrino propagation since the mass of neutrinos is different. This results in periodically changing detection probability of certain flavor, a phenomenon is called neutrino flavor oscillations.

We can study only disappearance effects with reactor neutrinos due to their energy range. The oscillation probability is in that case:
\begin{equation}
\begin{split}
P_{\bar\nu_e\rightarrow\bar\nu_e}(L,E)=1&-\sin^22\theta_{13}\left[\cos^2\theta_{12}\sin^2\left(\cos^2\theta_{12}\frac{\Delta m_{31}^2L}{4E}\right)+\sin^2\theta_{12}\sin^2\left(\frac{\Delta m_{32}^2L}{4E}\right)\right] \\
&-\sin^22\theta_{12}\cos^4\theta_{13}\sin^2\left(\frac{\Delta m_{21}^2L}{4E}\right)
\end{split}
\label{eq:prob}
\end{equation}
where $E$ is antineutrino energy, $L$ is the propagation distance, $\Delta m^2_{ij}=m^2_i-m^2_j$ are mass squared differences, from which two out of three are independent. Due to different values of mass squared differences, $\Delta m_{21}^2$ being $\sim$30 times smaller than $\Delta m_{31}^2$, the oscillation probability has two modes of oscillations. Oscillations at long baselines $\mathcal{O}(10\text{ km})$ are driven by $\Delta m_{21}^2$ and were explored by the KamLAND experiment \cite{Kamland}. Oscillations at short baselines $\mathcal{O}(1\text{ km})$ are driven by $\Delta m_{31}^2$ with an amplitude proportional to the $\theta_{13}$ mixing angle, which remained unknown until 2012, when the Daya Bay experiment proved a non-zero value of $\theta_{13}$ with more than $5\sigma$ \cite{dyb}.

We should note that the probability in Eq.~\ref{eq:prob} is derived within the three-neutrino framework. The reactor antineutrino anomaly, discussed later, together with several other measurements, could be explained by the existence of an additional neutrino with mass squared difference $\Delta m^2_{new}\simeq1\text{ eV}^2$. Additional mass eigenstates would result in new flavor states, which would need to be sterile, i.e. without any interaction of the Standard Model, due to the limit on number of light active neutrinos obtained by experiments at LEP collider \cite{LEP}. The search for a sterile neutrino is one of the main priorities in the field nowadays, extending beyond reactor neutrinos.

\subsection{Measurement of $\theta_{13}$ Mixing Angle and Mass Squared Difference $\Delta m_{32}^2$}
\begin{figure}[htb]
\begin{center}
\includegraphics[width=0.525\textwidth]{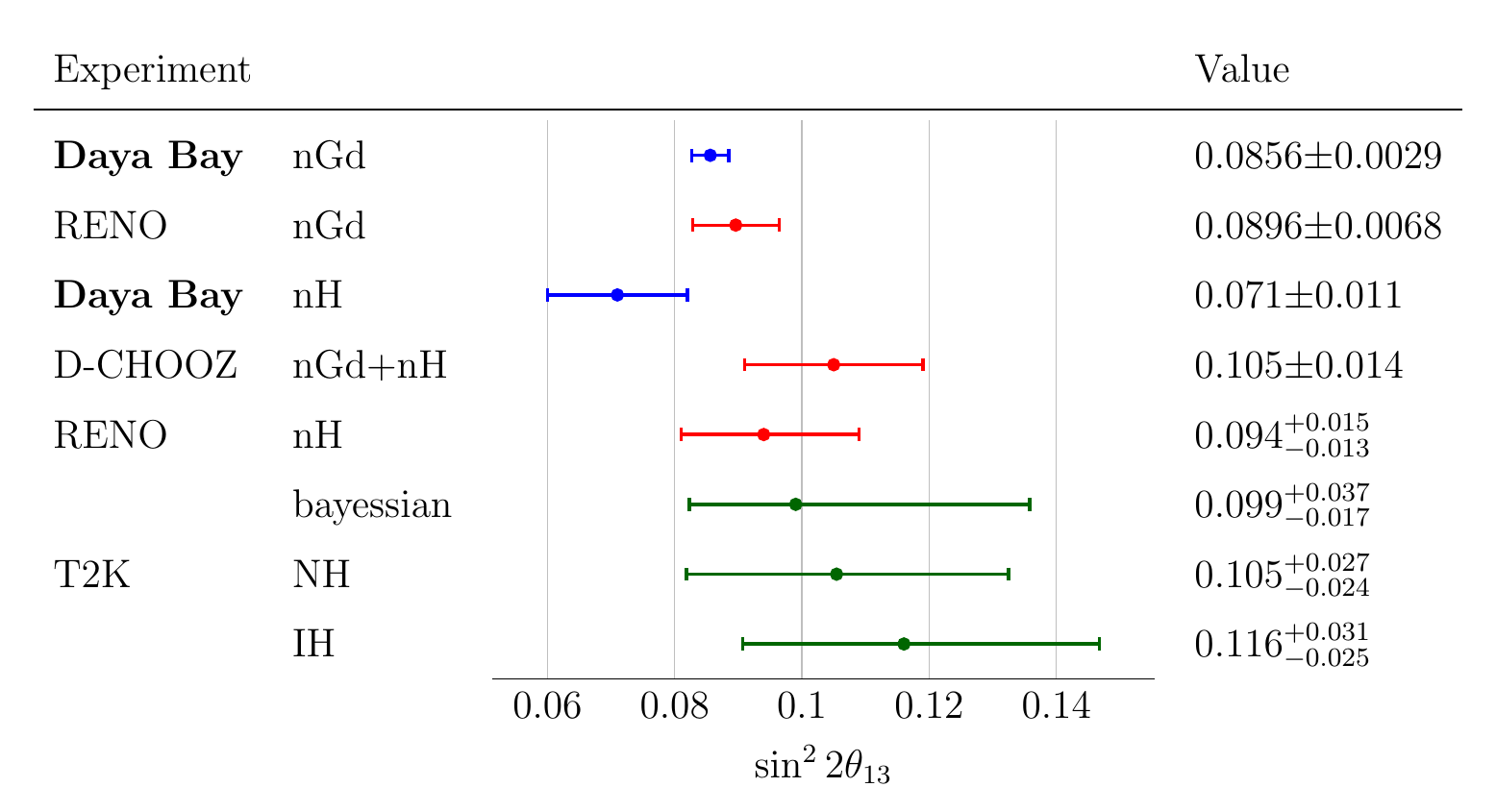}
\includegraphics[width=0.455\textwidth]{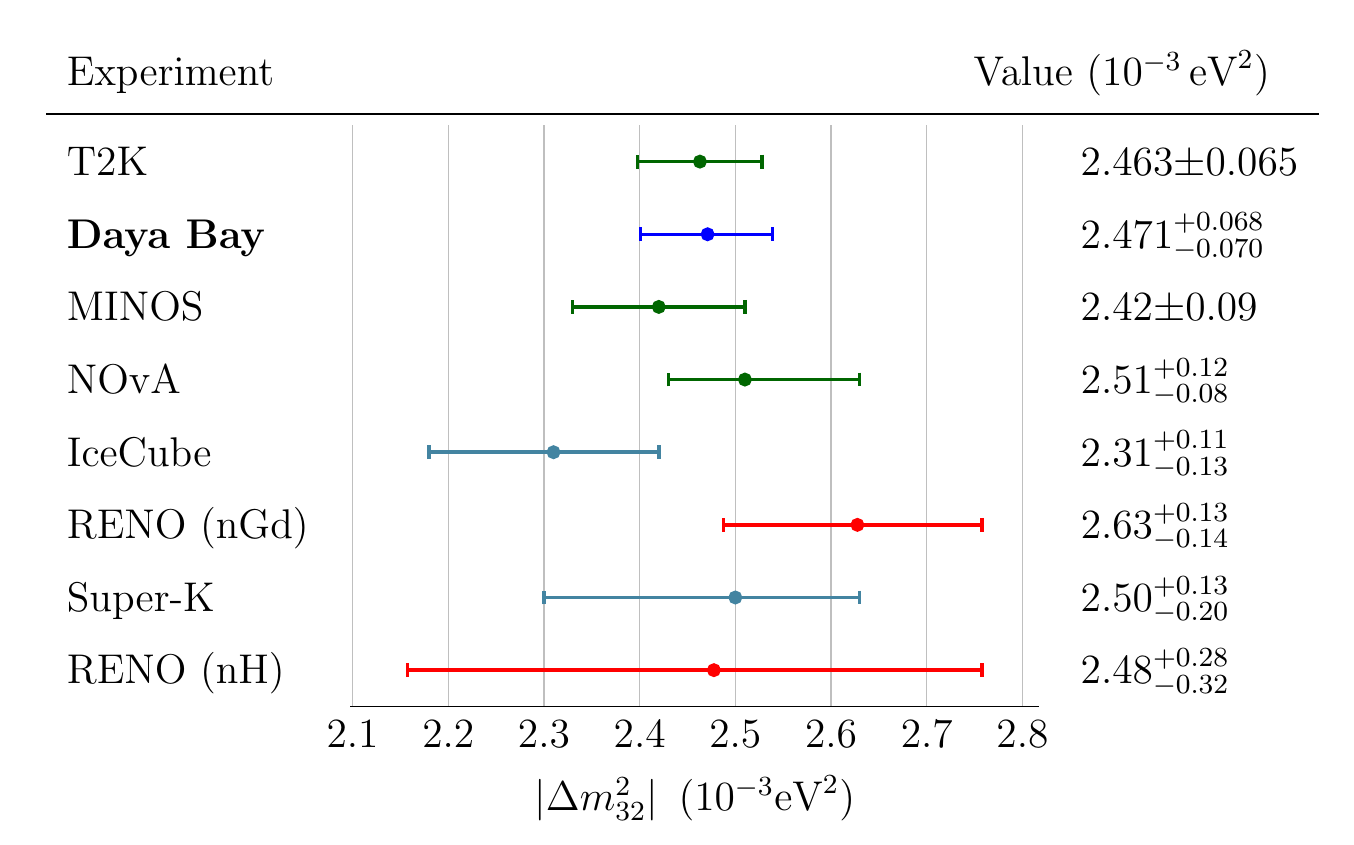}
\caption{\label{fig:best_osc} The latest measurements of the $\theta_{13}$ mixing angle \cite{DYB_latest,reno_latest,dyb_nh,bump_DC,reno_neutrino,t2k_angle} and mass squared difference $\Delta m^2_{32}$ \cite{t2k_latest,DYB_latest,minos_neutrino,nova_neutrino,ic_neutrino,reno_latest,sk_neutrino,reno_neutrino}, where the normal neutrino mass hierarchy is assumed.}
\end{center}
\end{figure}

Since 2012, the Daya Bay, RENO and Double Chooz experiments have reported gradually improved measurement of the $\theta_{13}$ mixing angle. The first two have also measured $\Delta m_{32}^2$. A summary of the latest measurements is shown of Fig.~\ref{fig:best_osc}.

Using a sample of antineutrinos selected with neutron capture on gadolinium, Daya Bay provides the world?s most precise measurement of $\sin^22\theta_{13}$, as well as a measurement of $\Delta m_{32}^2$ that is consistent and of comparable precision to that obtained from accelerator experiments. Double Chooz combines their data sets obtained with neutron capture on gadolinium, hydrogen and carbon in order to increase the statistics.

\section{Reactor Neutrino Anomalies}
In recent years, we have observed anomalies in the measurement of the absolute reactor neutrino flux as well as its energy spectrum. The difference comes from a comparison of the prediction model and the measurements, which are consistent among each other. We discuss the current state of anomalies.
\subsection{Reactor Neutrino Flux Anomaly}
\begin{figure}[htb]
\centering
\includegraphics[width=0.5\textwidth]{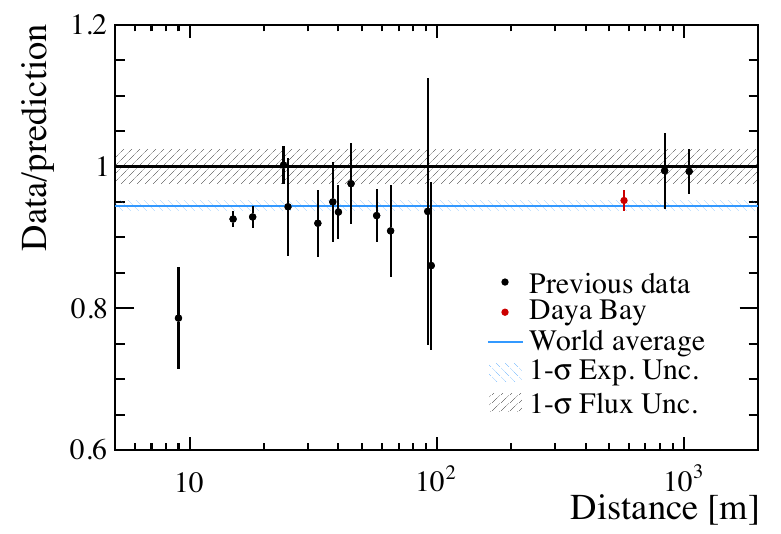}
\caption{The ratio of observation over the Huber+Mueller model prediction of past measurements corrected for the current knowledge of three-neutrino oscillations. The Daya Bay measurement of $R=0.952\pm0.014\text{ (experimental)}$ \cite{DYB_flux} is highlighted. The band around unity represents 2.3\% Huber+Mueller model uncertainty. The global average $R=0.945\pm0.007\text{ (experimental)}\pm0.023\text{ (model)}$ \cite{DYB_flux} indicates a deficit in the measured flux.}.
\label{fig:react_anomaly}
\end{figure}

Several experiments have consistently measured a reactor neutrino flux at distances $\mathcal{O}(10-100\text{ m})$ with about a 5\% deficit compared to the Huber+Mueller model \cite{Huber,Mueller}. The situation is graphically illustrated in Figure~\ref{fig:react_anomaly}, where the ratios of measurements over the prediction corrected with out best knowledge of three-neutrino oscillations are shown. A world average ratio $R=0.947\pm0.007\text{ (experimental)}\pm0.023\text{ (model)}$ \cite{DYB_flux} is obtained. 

The deficit can be explained by sterile neutrino oscillations. Even though sterile neutrinos do not interact weakly, they can still take part in oscillations. The presence of an additional flavor state would imply existence of additional mass state with the mass squared difference $\Delta m^2_{new}\gtrsim1\text{ eV}^2$ to form fast neutrino oscillations with length $\sim$$\mathcal{O}(\text{m})$. The sterile neutrino hypothesis is supported by several other discrepancies observed by LSND \cite{LSND}, MiniBooNE \cite{MiniBooNE}, SAGE \cite{SAGE} and GALLEX \cite{GALLEX}. 

More conservatively, issues with the prediction model can be behind the reactor neutrino anomaly. The final resolution whether there are sterile neutrinos behind flux anomaly would be observation of typical oscillation $\frac{L}{E}$ dependence pattern in the very short baselines. New precise experiments located close to the antineutrino source are coming online to tackle this matter. 

\subsection{Reactor Neutrino Spectrum Anomaly}
\begin{figure}[htb]
\begin{center}
\includegraphics[width=0.42\textwidth]{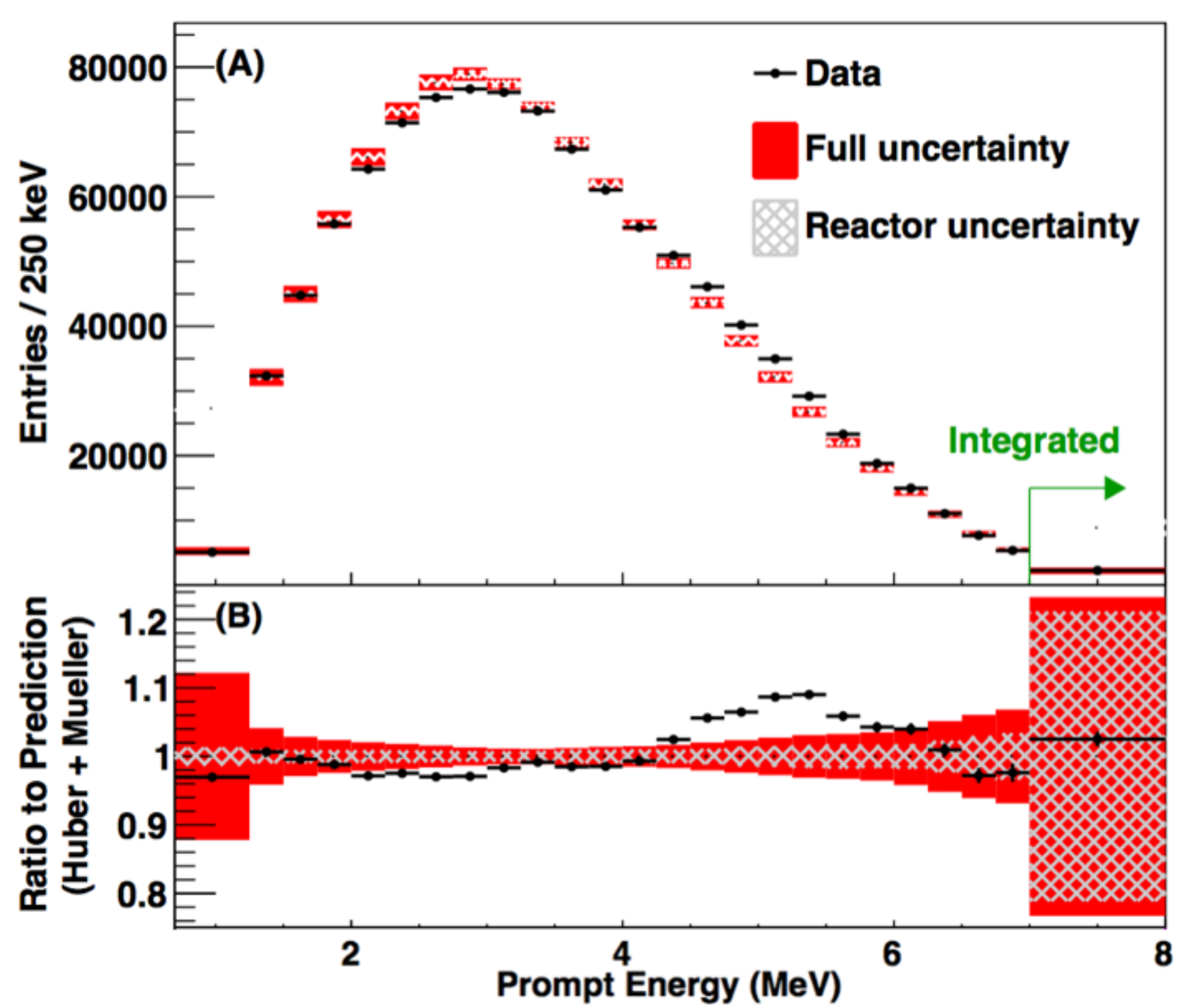}
\includegraphics[width=0.5\textwidth]{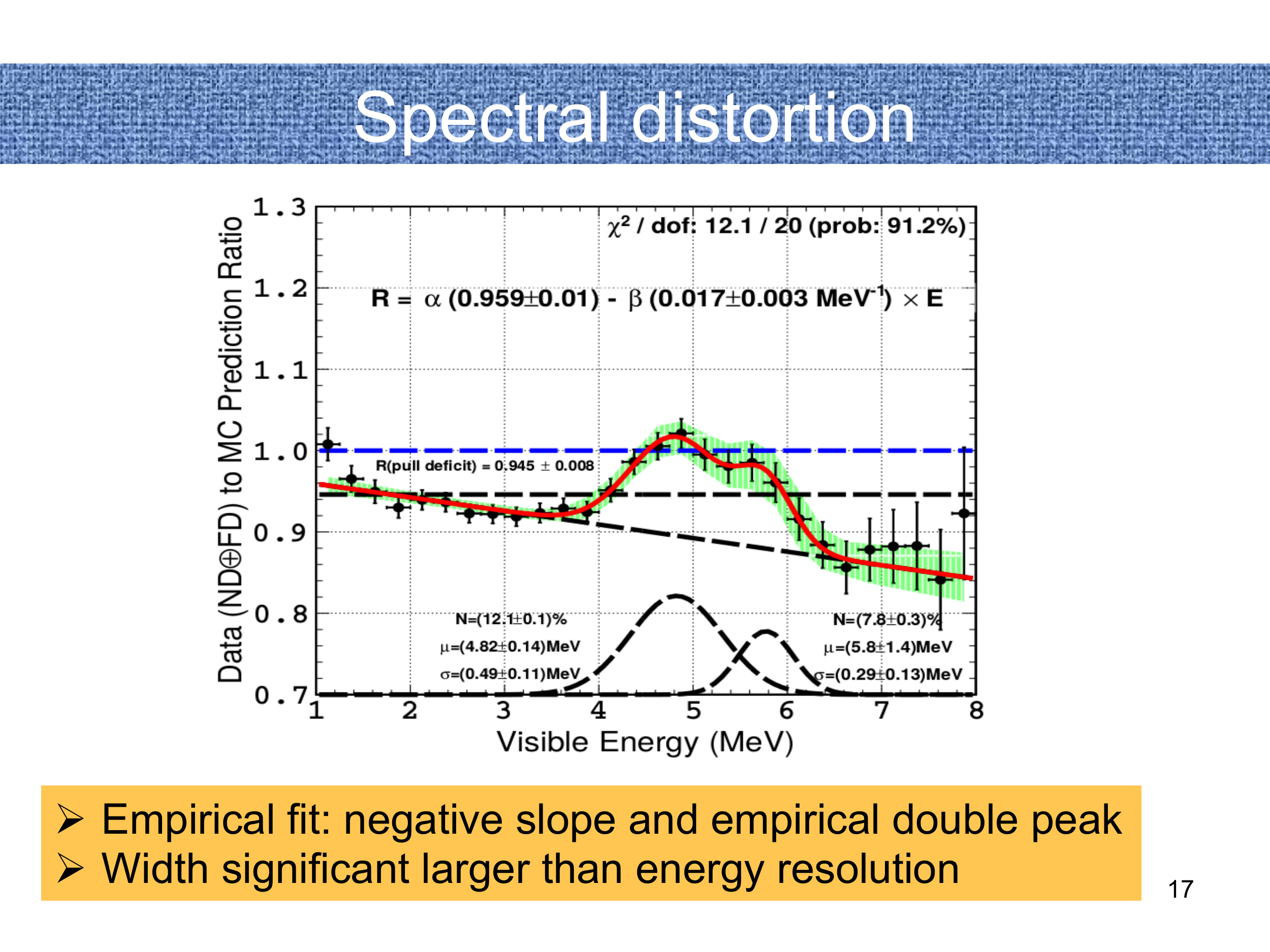}
\caption{\label{fig:bump} Examples of the excess of the measured reactor neutrino spectrum over the prediction observed by the Daya Bay experiment \cite{DYB_spectrum} (left) and by the Double Chooz experiment \cite{bump_DC} (right). The Double Chooz measurement suggests further structure in the bump, so far empirical.}
\end{center}
\end{figure}

Current experiments can test the reactor neutrino energy spectrum shape with unprecedented precision. An excess of detected antineutrinos over the prediction is observed in the energy range of $E_{\bar\nu_e}=5-7\text{ MeV}$. Such discrepancy was reported by several experiments \cite{DYB_spectrum, bump_DC, RENO_bump, NEOS_bump} and two examples are shown on Figure~\ref{fig:bump}. The most precise measurement comes from Daya Bay mainly due to the large collected statistics. The local significance of the `bump' is at the level of 4.4$\sigma$~\cite{DYB_spectrum}. The fact that the excess is correlated with reactor power \cite{RENO_bump} suggests an inaccurate prediction model. The spectrum shape anomaly cannot be explained neither by sterile neutrino oscillations, those would affect the whole spectrum, nor by a systematic bias related to the detector since it was not observed in other types of events such as the spallation $^{12}$B spectrum \cite{DYB_spectrum}.  The exact source of the `bump' remains unknown but several hypotheses have been proposed. It might come from $^{238}$U which is currently loosely constrained, the forbidden decay corrections are not as assumed or there is an intrinsic problem with the electron spectra used in the conversion \cite{ILL1, ILL2, ILL3} such as the hardness of the neutrons used in the measurement. A partial answer can be provided by HEU reactor experiments where only $^{235}$U is used as a nuclear fuel.

\subsection{Reactor Neutrino Flux Anomaly and Fuel Evolution }
\begin{figure}[htb]
\centering
\includegraphics[width=0.54\textwidth]{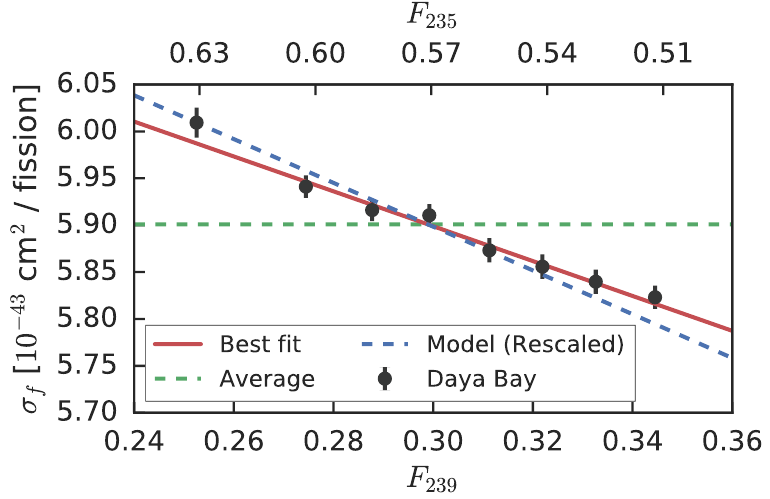}
\includegraphics[width=0.45\textwidth]{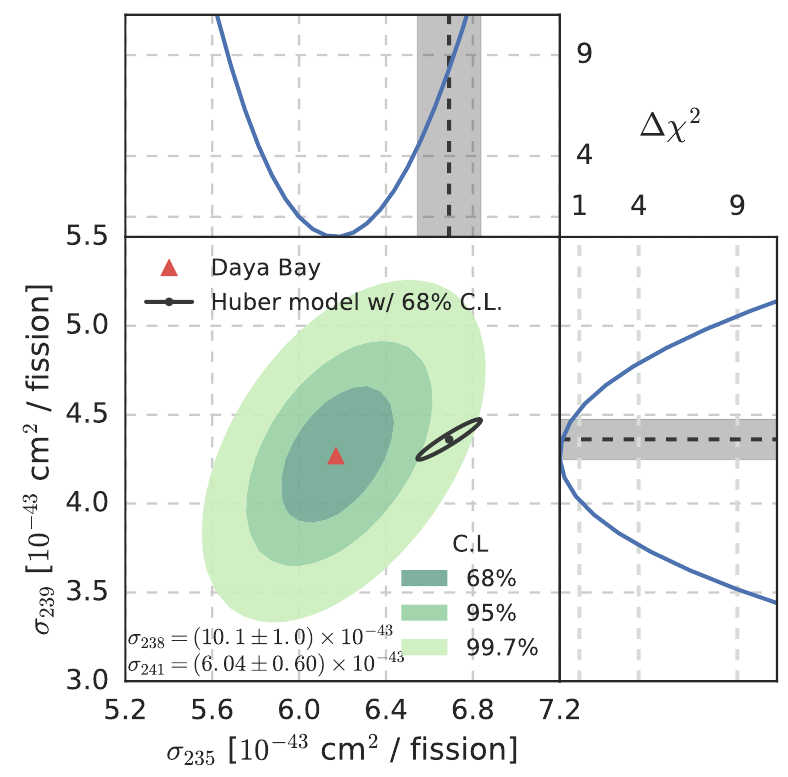}
\caption{The evolution of reactor neutrino flux expressed as IBD yield per fission as a function of effective $^{239}$Pu fission fraction \cite{DYB_evolution} (left). The slope of Daya Bay measurement (red line) does not match the predicted evolution of Huber+Mueller model (blue curve), which was corrected for the reactor antineutrino anomaly effect. The Daya Bay measured IBD yield per fission of $^{235}$U and  $^{239}$Pu \cite{DYB_evolution} and the comparison with Huber+Mueller prediction model \cite{Huber, Mueller} (right). The measurement is clearly lower than the prediction for $^{235}$U while $^{239}$Pu is in good agreement.}
\label{fig:evo}
\end{figure}

The relative contribution of four main isotopes to the number of fissions in the reactor changes during fuel cycle in LEU nuclear reactors. Most of the fissions come from $^{235}$U in the beginning while $^{239}$Pu builds up and is the main source at the end. Two other isotopes have a small and rather constant contribution throughout the cycle in pressurized (light) water reactors. We expect a change in the overall antineutrino flux with changes of fuel content since the antineutrino yield per fission varies slightly for the four isotopes. Daya Bay reported a measurement of such reactor neutrino flux and spectrum evolution \cite{DYB_evolution} and revealed a tension between the measurement and the evolution prediction.  Although the spectrum shape evolution agrees within current experimental precision, the evolution of the flux exhibits a different trend, as shown on the left panel of Figure~\ref{fig:evo}. Daya Bay further disentangled the yield of $^{235}$U and $^{239}$Pu, applying conservative constraints for the other two isotopes. A deficit between the predicted and measured IBD yield per fission was found for $^{235}$U while $^{239}$Pu agreed very well  with the model as shown on the right panel of Figure~\ref{fig:evo}. If sterile neutrinos were the sole cause for the the reactor anomaly, the same deficit should be observed in IBD yield for all isotopes. Since $^{235}$U has a significantly lower deficit then $^{239}$Pu compare to the prediction, an equal-deficit hypothesis was disfavored by Daya Bay at 2.8$\sigma$ weakening the sterile neutrino interpretation. However, Daya Bay did not rule out sterile neutrinos completely. Further investigation with current and upcoming experiments are needed to provide a final explanation for the reactor antineutrino anomaly.

\section{Reactor Neutrino Experiment Prospects and Conclusions}
Reactor neutrino experiments currently are also at the front line for addressing two pending questions in neutrino physics: neutrino mass hierarchy and possible existence of sterile neutrinos. 

The JUNO experiment currently under construction in China plans to measure reactor neutrino oscillations at the baseline of $\sim$52.5~km \cite{juno_yb}. JUNO will consist of a spherical 20~kt liquid scintillator detector, the largest of its kind, with a superb energy resolution of $<$3\% at 1~MeV. It will be the first experiment to observe both $\Delta m^2_{21}$-driven $\Delta m^2_{32}$-driven oscillations and due to that it will be able to determine the neutrino mass hierarchy with $>$$3\sigma$ significance. This is indeed allowed by the large value of the $\theta_{13}$ mixing angle. JUNO will also measure the $\theta_{12}$, $\Delta m^2_{21}$, $\Delta m^2_{32}$oscillation parameters with $<$1\% precision. The experiment will also be a powerful observatory for neutrinos from many other sources in addition to nuclear reactors. There will be a rich program for neutrinos from other sources, such as geoneutrinos, supernova neutrinos etc.

The next generation of precise very short baseline experiments such as NEOS, PROSPECT, STEREO, DANSS \cite{NEOS_bump, PROSPECT, STEREO,DANSS} etc. starts to provide results. These experiments, which are located at $\mathcal{O}(10\text{ m})$ baselines, are searching for sterile neutrinos with $\Delta m^2_{new}\approx1\text{ eV}^2$. In addition, they will provide precise measurements of the reactor neutrino spectrum, which is particularly interesting for those located at HEU research reactors. These reactors use primarily $^{235}$U as a fuel, limiting the spectral prediction to a single isotope and allowing for a direct comparison with the prediction. We note, that Daya Bay experiment suggested $\sim$8\% lower flux than predicted for this isotope \cite{DYB_evolution}.

Reactor antineutrinos were, are and will be great tools for investigating neutrino properties. The era of precise measurements not only provided us the value of the $\theta_{13}$ mixing angle value but also revealed several deviations from our reactor neutrino prediction models: a deficit of the overall flux and particular flux of $^{235}$U as well and excess in the reactor antineutrino spectrum in the energy range of $5-7\text{ MeV}$. Upcoming experiments plan to address these anomalies while also focusing on on elementary neutrinos properties such as determination of neutrino mass hierarchy.

\end{document}